\documentclass[final]{raa06}  

\usepackage{graphicx,times}             %for PS/EPS graphics inclusion, new
\usepackage{natbib}
\usepackage{amssymb,amsmath}
\bibpunct{(}{)}{;}{a}{}{,}
\usepackage[colorlinks=true, citecolor=blue]{hyperref}%
\usepackage{orcidlink}

\usepackage{graphicx,kantlipsum,setspace}
\usepackage{caption}
\usepackage{graphics,epsf}
\usepackage{amsmath}                % American Mathematical Society package
\usepackage{amsfonts}               % American Mathematical Society fonts
\usepackage{amssymb}                % American Mathematical Society symbol
\usepackage{epsfig}                 % EPS figures
\usepackage{appendix}
\usepackage{graphicx}
\usepackage{float}
\usepackage{color}
\usepackage{multirow}
\usepackage{colortbl}
\usepackage[para,online,flushleft]{threeparttable}
\usepackage{xcolor}

\hypersetup{citecolor=blue, % color for \cite{...} links
            linkcolor=red, % color for \ref{...} links
            menucolor=blue, % color for Acrobat menu buttons
            urlcolor=blue}  % color for \url{...} links
\newcommand{\cm}{{~\rm cm}}

\newcommand{\km}{{~\rm km}}
\newcommand{\s}{{~\rm s}}

\newcommand{\G}{{~\rm G}}

\newcommand{\yr}{{~\rm yr}}

\newcommand{\pc}{{~\rm pc}}

\begin{document}

   \title{Type Ia supernovae interacting with a close circumstellar material (SNe Ia-CSM) are SNe Ia inside planetary nebulae (SNIPs)
%\,$^*$
%\footnotetext{$*$ Supported by the National Natural Science Foundation of China.}
}
%   \subtitle{I. Place Your Subtitle Here}

   \volnopage{Vol.0 (20xx) No.0, 000--000}      %%preserved for Editor. DOn't remove!
   \setcounter{page}{1}          %%starting page, preserved for Editor. DOn't remove!

%\author{Noam Soker\,\orcidlink{0000-0003-0375-8987}} 
   \author{Noam Soker, \orcidlink{0000-0003-0375-8987}
     % \inst{1}
    }
%% Here is an example of three authors come from different institutes.
%% For single author or all the authors from an institute, use "\inst{}" only

   \institute{Department of Physics, Technion - Israel Institute of Technology, Haifa, 3200003, Israel;   {\it   soker@technion.ac.il}\\
%% Please give the E-mail address of the author, to whom future correspondence and
%% offprint requests will be sent.
%        \and
%             Full institute address for the third author\\
\vs\no
   {\small Received~~20xx month day; accepted~~20xx~~month day}}

\abstract{
I show that a newly estimated fraction of normal type Ia supernovae (SNe Ia) that interact within about 100 days of explosion with circumstellar material (CSM), called SNe Ia-CSM, is compatible with a recently estimated fraction of normal SNe Ia that interact with an old planetary nebula, hence, supporting the core-degenerate (CD) scenario for normal SNe Ia. According to the CD scenario, a white dwarf (WD) merges with the core of an asymptotic giant branch star at the end of common envelope evolution (CEE) and forms a massive WD remnant close to the Chandrasekhar mass. The CEE ejects a planetary nebula that the WD remnant ionizes. Most explosions occur within a merger-to-explosion delay (MED) time of less than a million years, before the planetary nebula material disperses to the interstellar medium, leading to a SN Ia inside a planetary nebula (SNIP). I discuss two plausible MED time distributions and show that the newly determined SNe Ia-CSM fraction of all normal SNe Ia, $\simeq 0.04 \%$, is compatible with the SNIP fraction of $\simeq 80\%$. Therefore, although the fraction of SNe Ia-CSM is very small, it does not require a rare evolutionary pathway. I argue that SNe Ia-CSM follow the same scenario that accounts for $70\% - 90\%$ of all normal SNe Ia, namely, the CD scenario. 
 }

 \authorrunning{N. Soker}            
\titlerunning{SNe Ia inside planetary nebulae}  
   
      \maketitle
% ==========================================================
\section{INTRODUCTION}
\label{sec:intro}
% ==========================================================
 
Despite the iconic Tycho and Kepler type Ia supernovae (SNe Ia) and the intensive usage of the images of their remnants, as well as other SN Ia remnants (SNRs), in public lectures and classes, there is no consensus on the leading scenarios of normal SNe Ia. Not only there is no consensus on the leading SN Ia scenarios, there is also no consensus on the classification of the different theoretical scenarios, as evidence from the many reviews in the last decade (e.g., \citealt{Maozetal2014, MaedaTerada2016, Hoeflich2017, LivioMazzali2018, Soker2018Rev, Soker2019Rev, Soker2024Rev, Wang2018,  Jhaetal2019NatAs, RuizLapuente2019, Ruiter2020, Aleoetal2023, Liuetal2023Rev, Vinkoetal2023, RuiterSeitenzahl2025}). 
A prominent piece of evidence from recent classifications and studies is that the previous millennium's classification of the single-degenerate (SD) and double-degenerate (DD) scenarios is obsolete. The core-degenerate (CD) scenario, studied in this paper, does not belong to either the SD or the DD scenarios.  
  
Some scenarios have two or more channels (sub-scenarios). Channels of a scenario differ in secondary parameters, like the composition of one or two WDs in the DD scenario or the evolutionary phase at which the WD accretes mass in the SD scenario. Examples include the common-envelope wind model \citep{MengPodsiadlowski2017, CuiMeng2022, WangMeng2025} of the SD scenario, the HeCO hybrid channel of the DD scenario (e.g., \citealt{Peretsetal2019, Zenatietal2019, Zenatietal2023}), the channels of the SD scenario with a main-sequence star and a carbon–oxygen–neon white dwarf (WD; e.g., \citealt{GuoMeng2025RAA}), and the core-merger detonation model \citep{Ablimit2021}. Some papers reconsider new processes, i.e., new mass-transfer prescriptions (e.g., \citealt{Lietal2023RAA}).
Each scenario has advantages, but also struggles to explain certain observations. Some observations challenge most scenarios, or even all (e.g., \citealt{Pearsonetal2024, SchinasiLembergKushnir2025, sharonKushnirWygoda2025, Sharonetal2025, Wangetal2024}).   

Because there is no one emerging leading SN Ia scenario, studies in recent years have considered the different scenarios and their channels for normal and peculiar SNe Ia (e.g., \citealt{Boraetal2024, Bossetal2024, Bregmanetal2024, CasabonaFisher2024, DerKacyetal2024, Joshietal2024, Koetal2024, Kobashietal2024, Limetal2024, Mandaletal2024, MandalSetal2025Tycho, Mehtaetal2024, Palicioetal2024, Phillipsetal2024, Soker2024RAAPN, Uchidaetal2024, Burmesteretal2025, Ferrandetal2025, Gabaetal2025NewA, Griffithetal2025, Hoogendametal2025a, Hoogendametal2025b, Itoetal2025, IwataMaeda2025, Kumaretal2025, Mageeetal2025, Michaelisetal2025, MichaelisPerets2025, OHoraetal2025,   Simotasetal2025, WangChenPan2025, ChenCetal2026, Courtetal2026, Dasetal2026, Pakmoretal2026, Panetal2026, Pateletal2026, Pollinetal2026, ZhaoXetal2026}, for a limited list of papers since 2024). 
The situation might be complicated by the presence of multiple SN Ia populations (e.g., \citealt{Gabaetal2026}), particularly Chandrasekhar- and sub-Chandrasekhar-mass explosions (see the reviews listed above).   
  
In addition to general explorations and studies of different scenarios, there are debates about the scenarios for specific SNe Ia. While \cite{Dasetal2025NatAs} and \cite{Mandaletal2025SNR0509} argued for the double-detonation (DDet) scenario for SNR 0509-67.5, I argued for the CD scenario \citep{Soker2025SNR0509}. The dispute is whether the metal distribution, like calcium, sulfur, and iron, matches simulations of the DDet scenario, or is better compatible with a SN Ia inside planetary nebulae (SNIP) in the framework of the CD scenario.  \cite{MandalSetal2025Tycho} argued for the DDet scenario for Tycho SNR, while I argued for the CD scenario \citep{Soker2025Tycho}; \cite{DickelJones1985} already argued that Tycho is an SNIP. The two papers used different analyses to reach their different conclusions: \cite{MandalSetal2025Tycho} analyzed the sizes of small-scale turbulent substructures of
different elements, while \citep{Soker2025Tycho} analyzed the outer structure, which shows two opposite protrusions, i.e., ears. 
Specifically, I argue that both SNRs are SNIPs. 

The CD scenario predicts that a large fraction of SNe Ia are SNIPs. Earlier papers estimated that a large fraction of SNe Ia are SNIPs: 
$f_{\rm SNIP}(2015) > 0.2$ \citep{TsebrenkoSoker2015},  
$f_{\rm SNIP}(2019) \simeq 0.15 - 0.2$ \citep{Soker2019CEEDTD},
$f_{\rm SNIP}(2022) \simeq 0.5$ \citep{Soker2022RAA}, 
$f_{\rm SNIP}(2025) >  0.5$ \citep{Soker2025SNR0509}, and 
$f_{\rm SNIP}(2026) \simeq 0.7-0.9$ \citep{Soker2025Tycho}.  
The increase in the estimated value of $f_{\rm SNIP}$ over the years resulted from new observations of SN Ia remnants and analyzing them with a specific search for morphological features similar to those of planetary nebulae and for massive hydrogen-rich circumstellar material (CSM), like the case of SNR 0509-67.5 and Tycho mentioned above. 

The goal of this study is to incorporate a new estimate by \cite{Desaietal2026} of the fraction of SNe Ia that interact early, within $\simeq 100$~days, with CSM, termed SNe Ia-CSM, into the CD scenario and the estimated fraction of SNIPs. Before I conduct this task in Section \ref{sec:Fractions}, I update the status of the six main scenarios in the classification I adapt here (Section \ref{sec:Scenarios}). I summarize in Section \ref{sec:Summary}.

% ==========================================
\section{Scenarios of normal SN Ia in 2026}
\label{sec:Scenarios}
% ==========================================

In this section, I present my 2026 updated estimate of the contributions of different scenarios to normal SNe Ia. I will refer to the classification I have been using in the last two years \citep{Soker2024Rev, Soker2025SNR0509, BraudoSoker2024, BraudoSoker2025RAA}. I present the six scenarios in Table \ref{Tab:Table1}. As I emphasized in Section \ref{sec:intro}, there are other classifications by different studies, as there is no consensus on the classifications. The first five rows are as in previous studies, as they describe the basic properties of the scenarios. 
The sixth row in Table \ref{Tab:Table1} describes the tension (if it exists) between observations and the scenarios for them to account for a large fraction of SNe Ia, and the seventh (next to last) row lists the theoretical challenges that the different scenarios encounter.  The last row gives my estimate of the likelihood of each scenario to account for a fraction of $\gtrsim 10\%$ of SNe Ia. 
% TTTTTTTTTTTTTTTTTTTTTTTTTTTTTTTTTTTTTTTTTTTTT
% Table generated by Excel2LaTeX from sheet 'Sheet1'
\begin{table*}
%\tiny
\scriptsize
%\footnotesize
\begin{center}
  \caption{An SN Ia scenarios classification: Challenges in explaining a large fraction of normal SNe Ia}
    \begin{tabular}{| p{1.8cm} | p{2.4cm} | p{2.4cm}| p{2.0cm}| p{2.0cm} | p{2.0cm} | p{2.0cm} |}
\hline  % ----------------------------
\textbf{Group} & \multicolumn{2}{c|}{$N_{\rm exp}=1$: {{Lonely WD}}}  &  \multicolumn{4}{c|}{$N_{\rm exp}=2$}     \\  
\hline  % ----------------------------
%\textbf{Outcome} & \multicolumn{2}{c|}{{{Mostly normal SNe Ia}}}  &  \multicolumn{4}{c|}{{{Mostly peculiar SNe Ia}}}\\ 
%\hline  % ----------------------------
\textbf{{SN Ia Scenario}}  & {Core Degenerate }    & {Double Degenerate - MED} & {Double Degenerate} & {Double Detonation} & {Single Degenerate} & {WD-WD collision} \\
\hline  % ----------------------------
\textbf{{Name}} & CD & DD-MED & DD & DDet & SD-MED or SD & WWC\\
\hline  % ----------------------------
\textbf{MED time} & MED  & MED  & 0  &0  & MED or 0 & 0 \\
\hline  % ----------------------------
 {$\mathbf{[N_{\rm sur}, M, Ej]}$$^{[{\rm 2}]}$}
  & $[0,M_{\rm Ch},{\rm S}]$ 
  & $[0,M_{\rm Ch}, {\rm S}]$ 
  & $[0,$sub-$M_{\rm Ch},{\rm N}]$
  & $[1,$sub-$M_{\rm Ch},{\rm N}]$
  & $[1,M_{\rm Ch},{\rm S~or~N}]$  
  & $[0,$sub-$M_{\rm Ch},{\rm N}]$ \\
\hline  % ----------------------------
\textbf{{Observational incompatibilities}} & & (1) $f_{\rm SNIP} > 0.5$ & (1) $f_{\rm SNIP} > 0.5$ (2) Predicts too asymmetrical SNR; (3) Many SNe Ia are $M_{\rm Ch}$ & (1) $f_{\rm SNIP} > 0.5$; (2) Predicts too asymmetrical SNR; (3) Many SNe Ia are $M_{\rm Ch}$; (4) No surviving companions in SNRs & (1) No surviving companions in SNRs; (2) WDs in novae lose mass;  (3) No hydrogen in SNe Ia & (1) $f_{\rm SNIP} > 0.5$; (2) Predicts too asymmetrical SNR; (3) Many SNe Ia are $M_{\rm Ch}$  \\
\hline  % ----------------------------
\textbf{{Theoretical challenges}} & (1) $M_{\rm WD} \simeq M_{\rm Ch}$; (2) The MED time and its parameters & (1) $M_{\rm WD} \simeq M_{\rm Ch}$; (2) The MED time and its parameters  & & & (1) Increase the $M_{\rm WD}$ to $M_{\rm Ch}$ & (1) Increase the collision rate \\
\hline  % ----------------------------
\textbf{{Contribution to normal SNe Ia }} & Most Likely $\qquad$ ($\simeq 70-90 \%$)& Likely  $\qquad  \qquad \qquad$  ($\simeq 10-30 \%$)& Possible $\qquad  \qquad \qquad$  ($<10\%$)& Less possible  & Highly unlikely & Highly unlikely  \\
\hline  % ----------------------------
     \end{tabular}
  \label{Tab:Table1}\\
\end{center}

\begin{flushleft}
\small 
Notes: An SN Ia scenarios classification scheme adapted from  \cite{Soker2024Rev}. The DDet channels with two exploding WDs, i.e., the DDet34 scenarios, are grouped with the DD scenario in this table, as these scenarios leave no surviving companion. The last three rows are new to the table from this study. The challenges in rows 6 and 7 refer to the possibility of the scenario accounting for a large fraction of SNe Ia ($\gtrsim 10 \%$). The last row is my estimate of the likelihood that the scenario contributes more than $10\%$ of normal SNe Ia. 
\newline
 Abbreviation. MED time: Merger to explosion delay time or mass transfer to explosion delay time.    
$N_{\rm exp}$: system's number of stars at the explosion moment. $N_{\rm sur}$: if a companion survives the explosion, then $N_{\rm sur}=1$, while if no star survives in the system $N_{\rm sur}=0$. In some peculiar SNe Ia, the exploding WD survives, and the system can have $N_{\rm sur}=2$. $M_{\rm Ch}$ and sub-$M_{\rm Ch}$ mark a near-Chandrasekhar-mass and sub-Chandrasekhar mass explosions, respectively. Ej is the ejecta morphology: S indicates scenarios that can lead to a spherical SNR, while N indicates scenarios that are expected to form SNRs with large departures from sphericity. 
\end{flushleft}
\end{table*}
% TTTTTTTTTTTTTTTTTTTTTTTTTTTTTTTTTTTTTTTTTTTTT

I will describe the scenarios from least to most likely (right to left in Table \ref{Tab:Table1}) and refer to some of the properties, challenges, and difficulties listed in the table; the rest are clear from the table and its caption.   

I first explain the tension between the four scenarios with the observations I will discuss in Section \ref{sec:Fractions}, namely, the claim that $f_{\rm SNIP} > 0.5$ (Section \ref{sec:intro} and  \citealt{Soker2025Tycho}). The planetary nebula that an asymptotic giant branch (AGB) progenitor of a WD expels disperses into the interstellar medium within $\approx 10^6 \yr$. Therefore, for the ejecta of an SN Ia to collide with an old planetary nebula (a SNIP), the explosion should occur within $\approx 10^6 \yr$ after the formation of the second WD in scenarios with two WDs. When two WDs come close together due to gravitational waves, the typical time will be much longer. Therefore, scenarios involving the merger of two WDs or their spiraling into proximity will not account for a large fraction of normal SNe Ia being SNIPS.  The crude estimate of the PN shell lifetime for interaction with the SN ejecta comes from the largest PNe. For example, the PN Sh 2-216 has a radius of $\simeq 1.7 \pc$ (e.g., \citealt{Tweedyetal1995}), and an expansion velocity of $<4 \km \s^{-1} $(\citealt{Reynolds1985}). This PN is still observed and intact. When it reaches a radius of $\simeq 10^{19} \cm$, twice as large as its present age, its age would be $\approx  8 \times 10^5 \yr$. This is the source of the estimated $10^6 \yr$.   

\textit{The WD-WD collision (WWC) scenario.} Studies show that the WWC scenario, where two unbound WDs collide and ignite (e.g., \citealt{Glanzetal2025b} for a recent study),  contributes much less than one per cent to normal SNe Ia (e.g., \citealt{Toonenetal2018}). I therefore consider this scenario's contribution to a significant fraction of normal SNe Ia highly unlikely (see the last column of Table \ref{Tab:Table1}). Also, the probability for an old PN is low, as the dynamical interaction that leads to a collision typically takes a very long time $ \gg 10^6 \yr$.

\textit{The single-degenerate (SD) scenario.} Recent observations, particularly by JWST, indicate the common presence of stable nickel, favoring near-Chandrasekhar mass WD: $M_{\rm WD} \simeq M_{\rm Ch}$ (e.g., \citealt{Hoeflichetal2026, KwokLetal2026}). It does not mean that there are no sub-$M_{\rm Ch}$ explosions, but that near-Chandrasekhar ($M_{\rm Ch}$) explosions are very common; these are the SD, DD-MED, and CD scenarios. 
The SD scenario, however, encounters severe problems that practically rule it out as a major contributor to normal SNe Ia. The challenge that observations do not detect surviving companions in SNRs Ia has been known for over a decade (e.g., \citealt{Kerzendorfetal2013, Kerzendorfetal2018}). The undetectable hydrogen that the ejecta is expected to strip from some donors in the SD scenario is another property that is incompatible with observations. 
The strong theoretical basis of the SD scenario was the idea that WD masses might grow in cataclysmic variables. However, later it was theoretically determined that this can occur only under certain conditions (e.g., \citealt{Hillmanetal2020, Vathachiraetal2024}), like a strong magnetic field of the WD (\citealt{AblimitMaeda2019}). Accretion of a helium-rich material might lead to a sub-$M_{\rm Ch}$ under specific conditions (e.g., \citealt{Hillmanetal2026}). A series of papers describing observations of different types of novae concludes that, despite being popular SN Ia candidates, the WD masses of these novae decrease, and, therefore, these cannot evolve to SNe Ia (e.g., \citealt{Schaefer2025a, Schaefer2025b, Schaefer2025c, Schaefer2026, SchaeferMyers2025, Schaeferetal2026}); a conclusion supported by some theoretical studies (e.g., \citealt{Vathachiraetal2026}). 
The SD scenario can supply CSM. However, it cannot be massive, as it is limited to a wind from an AGB star. Also, it cannot account for a large fraction of SNIPs, as most novae, which are the progenitors of (most) SNe Ia in the SD scenario, occur much later than $10^6 \yr$ after the formation of the WD. 
I consider all these observations to rule out the SD scenario as a contributor to normal SNe Ia. It might lead to some peculiar SNe Ia (which I do not study here). 
 
\textit{The double-detonation (DDet) scenario.} Many recent studies explore the properties of the DDet scenario (e.g., \citealt{MoranFraileetal2024, PadillaGonzalezetal2024, Polinetal2024, Shenetal2024, Zingaleetal2024, Callanetal2025, Rajaveletal2025, Wuetal2025, Mehtaetal2026}). However, the DDet scenario encounters challenges in explaining normal SNe Ia (e.g., \citealt{Soker2024Comment, Soker2025SNR0509}); it is more relevant to peculiar SNe Ia (e.g., \citealt{Glanzetal2025NatAs, Glanzetal2026}) and to the formation of runaway WDs (e.g., \citealt{Glanzetal2026}). Some runaway WDs might result from type Iax SNe (e.g., \citealt{Igoshevetal2023}), which is a group of peculiar SNe Ia, rather than those from normal SNe Ia coming from the DDet scenario. The DDet scenario has a surviving hot companion and highly asymmetrical ejecta, both of which are incompatible with many SNRs Ia. This scenario might, in principle, have an old planetary nebula around it, particularly for a helium non-degenerate mass donor. But most papers in recent years have considered a helium WD as the mass donor, and bringing the two WDs close together requires a long time in most, but not all, systems. I estimate that the DDet scenario might, at best, contribute a small fraction of normal SNe Ia ($\ll 10\%$). It might be a significant contributor to peculiar SNe Ia. 

\textit{The double-degenerate (DD) scenario (including DDet34). }
Under the DD scenario, I include cases in which both WDs explode within a few dynamical time scales of the merger. These include the violent merger (e.g., \citealt{Bhatetal2026}) with its highly asymmetrical explosion (e.g., \citealt{Pakmoretal2011}), and the ignition of both WDs in the triple or quadruple detonation channels (e.g., \citealt{Mehtaetal2026}). In the triple detonation (DDet3), the first WD explodes in the DDet process, and the ejecta triggers an explosion of the entire helium WD donor, leading to a highly non-spherical explosion \citep{Papishetal2015}; in the quadruple detonation (DDet4) the first WD explodes in the DDet process, and the ejecta triggers the ignition of the helium outer layer of the second WD, that in turn explodes the CO interior of the second WD (these two channels are termed together DDet34). The DDet34 scenario (or sub-scenario) leaves no stellar remnant. So I group it with the DD scenario rather than with the DDet scenario, in which the mass-donor WD or non-degenerate star survives (see, e.g., \citealt{Mehtaetal2026} for a recent study of DDet34). The violent merger process where a small fraction of the mass of one WD, its core, survives intact \citep{Pakmoretal2026}, is a channel (sub-scenario) of the DD scenario.  Practically, in these scenarios, there are two WDs at the time of explosion.
The two interacting WDs in most of these systems might no longer have planetary nebulae around them and cannot account for a large fraction of SNIPs. 
Two other main observational incompatibilities of the DD scenario are the prediction of highly non-spherical explosions and the observations that many SNe Ia are from the $M_{\rm ch}$ explosion (as mentioned above). 

The next two scenarios belong to the group of lonely WD scenarios, where at the explosion there is only one star in the system, a WD of mass $M_{\rm WD} \simeq M_{\rm Ch}$, which is a remnant of a merger. 

\textit{The DD with merger to explosion delay (MED) time (DD-MED) scenario.} In the DD-MED scenario, the explosion occurs at tens of dynamical time and more, even much more, after the merger. Namely, at explosion there is only one star in the system, a lonely WD of mass $M_{\rm WD} \simeq M_{\rm Ch}$. The merger of the two WDs is mainly due to gravitational wave emission, implying a very long time delay from the common envelope evolution (CEE), $\gg 10^6 \yr$. By the time of explosion, the mass that was ejected in the CEE of the first WD inside the envelope of the AGB progenitor of the second WD is long dispersed in the interstellar medium.  In these cases, there will be no interaction of the SN Ia ejecta with an old planetary nebula. However, during the WD merger, some of the lighter WD's mass may be ejected. This is a carbon-oxygen-rich gas. In \cite{Soker2025DDMED} I suggested the DD-MED scenario with a MED time of $\simeq 1-2 \yr$ to account for the rare properties of SN Ia SN 2020aeuh, which are the interaction of the ejecta with a carbon–oxygen-rich CSM at $\simeq 50$~days post-explosion \citep{Tsalapatasetal2025}. I estimate that the DD-MED scenario contributes $\simeq 10-30 \%$ of normal SNe Ia. 

\textit{The core-degenerate (CD) scenario.} In the CD scenario, a WD merges with the core of an AGB star to form a WD of a mass $M_{\rm WD} \simeq M_{\rm Ch}$. The CEE triggers the merger rather than gravitational waves. The envelope that the CEE ejects forms a planetary nebula. After a MED time much larger than the dynamical timescale, the WD explodes. In most cases, the ejecta later collide with the planetary nebula (or former planetary nebula), i.e., a SNIP.  Recently, \cite{Macrieetal2026} reported the detection of $\approx 10^{-4} M_\odot$ of CSM dust at about one light year from the normal SN Ia  2023qov. I suggest that this dust is in an old planetary nebula; namely, SN 2023qov is a SNIP. The dust is at the inner boundary of the planetary nebula shell, and cooler dust might reside further out.  Because the WD merger with the core is not driven by gravitational waves, SNe Ia in the CD scenario occur much closer to star formation than in the DD and DD-MED scenarios. \cite{Heigeretal2026} claimed recently that $75\%$ of the SNe Ia occur within $\simeq 3.3 \times 10^8 \yr$ after a burst of star formation. The CD scenario is compatible with this large population of prompt SNe Ia.  The explosion of the $M_{\rm Ch}$ WD in the CD and DD-MED scenarios is the most spherical among all scenarios (e.g., \citealt{Soker2019Rev}); the $M_{\rm Ch}$ WD explosion in the SD scenario is also spherical, but the collision of the ejecta with the companion somewhat distorts the sphericity.      

Although the CD scenario is better compatible with observations than the other scenarios, it faces major theoretical challenges: (1) explaining the formation of an $M_{\rm Ch}$ WD in the core-WD merger process, and (2) the MED time until explosion. The DD-MED scenario faces the same theoretical challenges. 
Despite these challenges, the compatibility with observations should motivate the study of the merger process without an immediate explosion and the study of the MED time.
At present, the answers to these challenges are qualitative. The reasoning is as follows. (1) Because of the hot core, the merger of the CO-rich core with the CO WD does not lead to an immediate explosion (e.g., \citealt{KashiSoker2011}). The merger process occurs through a feedback mechanism: As the remnant approaches the Chandrasekhar mass, it contracts and releases gravitational energy, ejecting mass and preventing further accretion. (2) The young merger remnant is rapidly rotating. Only after it loses angular momentum does it reach the condition for ignition (e.g.,  \citealt{IlkovSoker2012, Neopaneetal2022}). 
 Section \ref{sec:Fractions} elaborates on the second point.

% ==========================================
\section{The fraction of SNIPs and SNe Ia-CSM}
\label{sec:Fractions}
% ==========================================

% ==========================================
\subsection{Uniform explosion probability}
\label{subsec:Uniform}
% ==========================================

\cite{TsebrenkoSoker2013} pointed out the potential of using SNIPs to reveal the SN Ia scenario, and \cite{TsebrenkoSoker2015} argued that a large fraction of SNe Ia are SNIPs; the latest estimate is that $f_{\rm SNIP}(2026) \simeq 0.7-0.9$ \citep{Soker2025Tycho}. I attribute immense importance to this ratio, because, as I argued in Section \ref{sec:Scenarios}, it suggests that the CD scenario is the primary contributor to normal SNe Ia. However, this estimate must always be checked against new observations. I here compare this estimate with the new finding by \cite{Desaietal2026} that SNe Ia-CSM with an ejecta-CSM interaction time of $\lesssim 100~{\rm day}\simeq 0.3 \yr$ after the explosion account for $0.036 \pm 0.017\%$ of the total local rate of all SNe Ia; they also find that normal SNe Ia are $\simeq 93\%$ of all SNe Ia. So the SNe Ia-CSM fraction out of normal SNe Ia is somewhat larger, 
\begin{equation} 
f_{\rm C03} \equiv f^{\rm obs}_{\rm CSM} (t_{\rm MED} \lesssim 0.3 \yr)\simeq (3.9 \pm 1.8)\times 10^{-4}. 
     \label{eq:fc03}
\end{equation} 
 The ratio of observationally inferred SNe Ia-CSM to observationally inferred SNIPs is then 
 \begin{equation}
\eta_{\rm obs} \equiv \left( \frac{f_{\rm C03}}{f_{\rm SNIP}}  \right)_{\rm obs} \simeq 4.9 \times 10^{-4} \pm 2.3 \times 10^{-4} .
    \label{eq:etaobs}
\end{equation}
To the accuracy of the following calculation I take $\eta_{\rm obs} \approx 5 \times 10^{-4}$. 
 The SNe Ia-CSM in the sample of \cite{Desaietal2026} suffered ejecta-CSM interaction within 100 days after explosion. This implies a radius of first interaction of $\lesssim 10^{16} \cm$  

The meaning of SNe Ia-CSM, and hence the fraction $f_{\rm C03}$, here are different than what I used in \cite{Soker2019CEEDTD}. Here, I refer to SNe Ia-CSM as the new study by \cite{Desaietal2026}, who consider interaction within $\simeq 0.3 \yr$ after explosion, while in \cite{Soker2019CEEDTD} I considered interaction within $t_{\rm MED} \simeq 3000 \yr$, namely, a much large interaction time hence large fraction $f_{\rm CSM}$. 

In the past, I assumed that the probability of exploding within a time interval $dt$ at MED time $t_{\rm MED}$ after the core-WD merger in the CD scenario is crudely constant up to a maximum MED time $t_{\rm SNIP}$. Namely, the majority (or even all) explosions occur within a MED time of $t_{\rm MED} \lesssim t_{\rm SNIP}$. This gives an explosion rate of    
\begin{equation}
\left( \frac{d N} {d t_{\rm MED}} \right)_{\rm con}  = \frac{N_{\rm SNIP}}{t_{\rm SNIP}} \qquad {\rm for} \qquad t_{\rm MED}<t_{\rm SNIP}, 
     \label{eq:ProbabilityConstant}
\end{equation}
where $N_{\rm SNIP}$ is the total number of SNIPs. 
In \cite{Soker2022RAA} I assumed $t_{\rm SNIP} \approx 10^4 \yr$, and in \cite{Soker2025SNR0509} I raised this number to $t_{\rm SNIP} \gtrsim 10^5 \yr$. 
This estimate is based on two recent identifications of SNR  0509-67.5 and Tycho as SNIPs, both with a radius of $R_{\rm SNR} \simeq 10^{19} \cm$. I scale the velocity of the inner boundary of the dense gas of the planetary nebulae, the slowest part of the nebula, with $v_{\rm PN} =10 \km \s^{-1}$. I also assume that the interaction of the ejecta with the old planetary nebula in these two SNRs started recently. These give a MED time of $R_{\rm SNR}/v_{\rm PN} \approx 3 \times 10^5 \yr$. 

Here, I consider that an ejecta-nebula interaction can start at radii of up to $\simeq 10^{19} \cm \simeq 3 \pc$ to form an SNIP. Consider a constant velocity of the front of the SN ejecta and an explosion rate as in equation (\ref{eq:ProbabilityConstant}), implies that $0.1\%$ of SNIPs will suffer an interaction within a radius of $r=10^{16} \cm$, implying an interaction within about 100 days for an ejecta velocity of $v_{\rm ej} \simeq 10^4 \km \s^{-1}$. This simple estimate, therefore, yields a theoretical value of 
\begin{equation}
    \eta_{\rm con} \simeq 10^{-3} .
    \label{eq:etaconstant}
\end{equation}  
This estimated value, based on equation (\ref{eq:ProbabilityConstant}), should be compared with the inferred ratio from observations as given by equation (\ref{eq:etaobs}). 
Considering the many uncertainties in the observational estimates (equation \ref{eq:etaobs}) and the simple constant explosion rate of equation (\ref{eq:ProbabilityConstant}), I consider the SN Ia-CSM fraction that \cite{Desaietal2026} found to be crudely compatible with the estimated fraction of SNIPs. 

\cite{Desaietal2026} argued that the fraction of SN Ia-CSM that they find implies that the evolutionary pathways producing massive CSM
shells around SN Ia progenitors are exceptionally
rare in the local universe. 
I argue, on the other hand, that SNe Ia-CSM do belong to the primary scenario, which I argue to be the CD scenario, but the exploding WDs of SNe Ia-CSM have a short MED time of $t_{\rm MED} \lesssim 300 \yr$. So the rare property is not the progenitor and scenario, but only the very short MED time.  
 
% ==========================================
\subsection{The role of the WD magnetic field}
\label{subsec:MagneticFields} 
% ==========================================

I consider an alternative to the constant explosion rate given by equation (\ref{eq:ProbabilityConstant}). 

I adopt the assumption that the merger product should lose angular momentum to explode (e.g., \citealt{IlkovSoker2012, Neopaneetal2022}), as was also suggested in the single degenerate scenario (e.g., \citealt{DiStefanoetal2011, Justham2011}). 

The mechanism for the WD merger remnant to lose angular momentum (e.g.,  \citealt{IlkovSoker2012, Neopaneetal2022}) is the magneto-dipole radiation torque, as used in pulsars (e.g., \citealt{Benacquistaetal2003, ContopoulosSpitkovsky2006}), which \citet{Benacquistaetal2003} also used for WDs.    
The spin-down time from an initial fast rotation ${\tilde{\Omega}_0}$ down to a critical angular velocity $\tilde{\Omega}_c$ at explosion occurs according to the assumption, is  
\begin{equation}
\begin{split}
 \tau_{\rm B} & \simeq
\frac{I c^3}{ B^2 R^6 \tilde\Omega_{\rm c}^{2}}
\left[1-\left(\frac{\tilde{\Omega}_0}{\tilde{\Omega}_c}\right)^{-2}\right] (\sin \delta)^{-2}
\approx 2 \times 10^{4}
\\&
\times
\left(\frac{B}{10^9 \G}\right)^{-2}
\left(\frac{\tilde\Omega_{\rm c}}{0.7 \Omega_{\rm Kep}}\right)^{-2}
\left(\frac{R}{4000 \km }\right)^{-1}
\\&
\times
\left(\frac{\sin \delta}{0.3}\right)^{-2}
\left(\frac{\beta_I}{0.3}\right)
\frac{\left[1-\left(\frac{\tilde{\Omega}_0}{\tilde{\Omega}_c}\right)^{-2}\right]}{0.2}
\yr.
\end{split}
\label{eq:taub}
\end{equation}
where $\delta$ is the angle between the magnetic axis and the rotation axis,
and $\beta_I$ is defined by $I = 0.4 \beta_I M_{\rm WD} R^2$, where $I$, is the moment of inertia of the WD and $R$ its radius. 
I scale ${\tilde{\Omega}_c}$ at which the WD explodes, as \cite{IlkovSoker2012} did, based on the results of \citet{YoonLanger2005} for WDs in the range of $1.4-1.5 M_\odot$. 
I assume that many WDs form with a rotation period not far from the critical one, as the critical $\tilde \Omega$ is $0.7$ times the maximum possible spin.  
For the magnetic field, I took a value between $10^{8} \G$ that \cite{IlkovSoker2012} scaled with, and $10^{10} \G$ that \cite{Neopaneetal2022} scaled with. 
The value of $\sin \delta$ can vary between zero and one; I scaled with $\sin \delta=0.3$, which is between the value of $0.1$ that \cite{IlkovSoker2012} scaled with, and $1$ that \cite{Neopaneetal2022} scaled with. For a random distribution of angles in all directions, the average value is $<\sin \delta>=\pi/4=0.785$. However, due to a dynamo operation that amplifies the magnetic field, we expect a tendency to alignment. Therefore, a scaling with $\sin \delta = 0.3$ is reasonable. 

The angular-momentum-loss timescale is sensitive to $\sin \delta$ and $B$.  
The value of $\sin \delta$ can differ from the scaling value by a factor of a few. On the other hand, the magnetic field can, in principle, vary by orders of magnitude. However, the merger of two WDs (e.g., \citealt{GarciaBerroetal2012}) or a WD with a core of an AGB star (e.g., \citealt{Briggsetal2018}) leads to a WD merger product with a strong magnetic field in a limited range. The calculations and comparison with observation by \cite{Briggsetal2018} suggest that the merger of a WD and the core of an AGB star can form WD remnants with magnetic fields of $10^8 \G \lesssim B \lesssim 10^9$; see, e.g., \cite{Malheiroetal2026} for a recent study of a massive WD which is a merger product with $B \simeq 10^9 \G$, and \cite{Kepleretal2013} for several WDs close to the Chandrasekhar mass and with strong magnetic fields. The magnetic fields are stronger for more compact merging WDs. Since I deal here with a core and a WD that add up to the Chandrasekhar mass or more, they are very compact. I take the magnetic field range at merger time to be on the upper end: $3 \times 10^8 \G \lesssim B \lesssim 3 \times 10^9 \G$.  
Such a strong magnetic field has implications not only for the MED time but also for the explosion process, particularly the transition from deflagration to detonation \citep{Shiberetal2026}. In the later process, convection and rotation during nuclear burning before runaway at explosion, i.e., during the smoldering phase, further amplify the magnetic field \citep{Shiberetal2026}. 

To derive a plausible MED time distribution, I write equation (\ref{eq:taub}) with reference only to the dependence on $\delta$ and $B$ and define $B_\delta \equiv B \sin \delta$. 
Namely, I take in this derivation $\tilde{\Omega}_0 = 1.1 \tilde{\Omega}_c$. As I indicated, I assume here that the merger product's initial rotation is not far from the critical one. Future study will have to include the dependence on $\tilde{\Omega}_0$, which might be significant. This will be done after accurate merger simulations yield the distribution of $\tilde{\Omega}_0$ in different core-WD mergers.   
The relation is then 
 \begin{equation}
 \tau_{\rm B} \approx 2 \times 10^{4}
\left(\frac{B_\delta}{ 3 \times 10^8 \G}\right)^{-2}  \yr =2 \times 10^{4} B^{-2}_{\delta 3} \yr ,
\label{eq:taubdelta}
\end{equation}
where 
\begin{equation} 
B_{\delta 3} \equiv \frac{ B \sin \delta}{3 \times 10^8 \G}. 
\label{eq:Bdelta3  d}
\end{equation}
The relevant range is approximately $3 \times 10^7 \G \lesssim B_\delta \lesssim 3 \times 10^9 \G$. The magnetic field relates to the time of the explosion by 
\begin{equation}
    B_{\delta 3} \approx 141 (t_{\rm MED}/\yr)^{-1/2} =  141 t^{-1/2}_1,  
    \label{eq:Bdelta3}
\end{equation} 
where the second equality defines 
\begin{equation}
    t_1 \equiv t_{\rm MED}/1 \yr.
    \label{eq:t1}
\end{equation}
The explosion rate is given by 
\begin{equation}
\frac {dN}{dt_{\rm MED}} =  \frac{d N}{dB_{\delta 3}} \left| \frac {dB_{\delta 3}}{dt_{\rm MED}}  \right| \approx 70 \frac{d N}{dB_{\delta 3}} t^{-3/2}_1 \yr^{-1},   
\label{eq:Rate1}
\end{equation}
where $N(B_{\delta 3})$ is the distribution of merger remnants as a function of $B_{\delta 3}$. 
Based on the results of \cite{Briggsetal2018}, a crude distribution would be a Gaussian on a log scale  
\begin{equation}
 \frac{d N}{d \log B_{\delta 3}} \approx N_0
 \exp\left\{ -\left( \frac{ \log B_{\delta 3}- \log B_{\delta 3,0}} {\Delta} \right)^2   \right\}  ,
\label{eq:Bdistribution1}
\end{equation}
 with a typical value of $\log B_{\delta 3,0} \simeq 0$. To have a very low number of systems at values an order of magnitude larger than $B_{\delta 3,0} =1$, the demand is that $\Delta \lesssim 1$. The value of $N_0$ has no meaning here, and can be set to $N_0=1$. The uncertainties are large. To reach the goal of this paper, which is to show that the new estimate of the fraction of SNe Ia-CSM is compatible with the estimated fraction of SNIPs, I will take a simple expression of 
 \begin{equation}
 \frac{d N}{d \log B_{\delta 3}} \approx N_0
 \exp \left\{ -\left( \log B_{\delta 3} \right)^2/\Delta^2  \right\}. 
\label{eq:Bdistribution2}
\end{equation}
Substituting equation (\ref{eq:Bdistribution2}) in equation (\ref{eq:Rate1}), using equation (\ref{eq:Bdelta3}) for $B_{\delta 3}$, yields
\begin{equation}
\frac {dN}{dt_{\rm MED}}  \approx 0.5 \frac{N_0}{t_1}  
\exp \left\{ - \frac{1}{\Delta^2}\left[ \log \left( \frac{141}{\sqrt{t_1}} \right) \right]^2  \right\}
\yr^{-1}.  
\label{eq:Rate2}
\end{equation}
 
I recall that $t_1$ is the MED time, namely, the time from the merger, which in the CD scenario is about the end of the CEE, to the explosion.
Equation (\ref{eq:Rate2}) presents a plausible general behavior of the rate of explosions, namely, the distribution of the MED time.  This expression for the explosion rate has a maximum at 
 \begin{equation}
t_{\rm MED}^{\rm max}(\Delta) \approx 2 \times 10^4 10^{-2\Delta^2 \ln 10} \yr. 
\label{eq:tmax1}
\end{equation}
The times of maximum rate of explosion from equation (\ref{eq:tmax1}) for three values of $\Delta$ are $t_{\rm max}(0.3) \approx 7.7 \times 10^3 \yr$, $t_{\rm max}(0.4) \approx 3.7 \times 10^3 \yr$, and $t_{\rm max}(0.5) \approx 1.4 \times 10^3 \yr$. These times of maximum rates for different values of $\Delta$ are for the parameters according to the scaling in equation (\ref{eq:taub}), and, therefore, are only crude estimates. 

\begin{figure*}
\begin{center}
%\vspace*{-5.9cm}
%\hspace*{-1.1cm}
\includegraphics[trim=0.0cm 0.0cm 0.0cm 0.0cm ,clip, scale=0.450]
{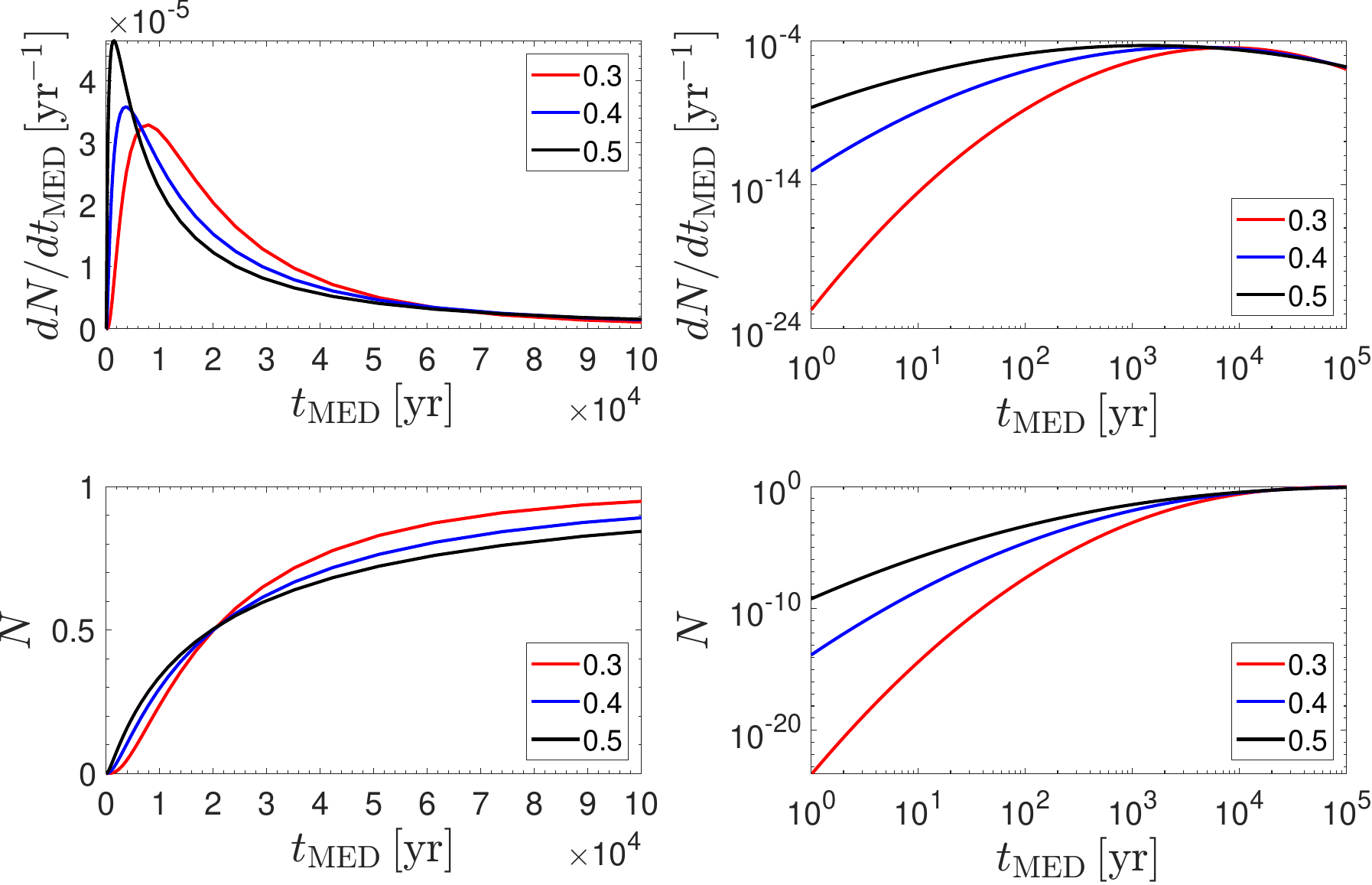}
%{TypeIaCSMSNIPfigLogRAA}
%\vspace*{-5.9cm}
\caption{The distribution (\ref{eq:Rate2}) in normalized systems per year (upper panels) and its integration from $t=0$ (lower panels), both as a function of the MED time in years. The normalization is for $N=1$ at $t=10^6 \yr$. The left panels use a linear scale, and the right panels use a log-log scale. 
}
\label{fig:Nmed}
\end{center}
\end{figure*}

In Figure \ref{fig:Nmed}, I present the distribution (\ref{eq:Rate2}) (upper panels) and its integration (lower panels); the left panels are on a linear scale, and the right panels are on a log-log scale. I normalized the number of systems to be $N=1$ at $t=10^6 \yr$, where the function in equation (\ref{eq:Rate2}) is very small.   
Equation 13 can be written in a different form by defining $z=\log(141/\sqrt{t_1})$, and so $d z = -(1/2)dt_1/t_1$, the limits of $t_1=0$ to $t_1=\infty$ become $z=+\infty$ to $z=-\infty$, respectively. Substituting these in the equation (\ref{eq:Rate2}) gives 
\begin{equation}
\frac {dN}{d z}  \approx {N_0}  \exp \left( - \frac {z^2}{\Delta ^2} \right) 
\yr^{-1}.  
\label{eq:Rate3}
\end{equation}
This function is symmetric about $z=0$, corresponding to $t_1=(141)^2$. Namely, $N=0.5$ at $t_{\rm MED}=2 \times 10^4 \yr$, as evident from Figure \ref{fig:Nmed}; half the systems have $t_{\rm MED} < 2 \times 10^4 \yr$. 
% %% FFFFFFFFFFFFFFFFFFFFFFFFFFFFFFFFFFFFFFFFFFFFFFFFFF 
% %% FFFFFFFFFFFFFFFFFFFFFFFFFFFFFFFFFFFFFFFFFFFFFFFFFF

The important point for this study is the relative fraction of SNe Ia-CSM to the fraction of SNIPs. Namely, the fraction of those with a MED time of $\lesssim 300 \yr$. For an SN ejecta $\simeq 1000$ times faster than the inner boundary of the planetary nebula, the interaction will take place within $\approx 0.3 \yr \simeq 100~{\rm day}$. Integrating equation (\ref{eq:Rate2}) from zero to $t_1=300$ and dividing by integration from zero to $3 \times 10^5\yr$, the time where interaction can still take place with a planetary nebula  (which is about the same as integrating to infinity, as there are not many systems with a MED time beyond that), gives the fraction of SNe Ia-CSM, $f_{\rm C03}$, to that of all SNIPs, $f_{\rm SNIP}$ (note that in the CD scenario, SNe Ia-CSM are a subgroup of SNIPs) as 
\begin{equation}
\eta_\Delta (\Delta) \equiv \left( \frac{f_{\rm C03}}{f_{\rm SNIP}} \right)_\Delta \approx 
\frac{ \int_0^{300 \yr} \frac{dN}{dt_{\rm MED}} dt_{\rm MED}}{\int_0^{3\times 10^5 \yr} \frac{dN}{dt_{\rm MED}} dt_{\rm MED}} . 
    \label{eq:etadeltaE}
\end{equation}
I find the distribution I assumed here to fit the observations (equation \ref{eq:etaobs}) for $\Delta =0.39$ with  
\begin{equation}
    \eta_\Delta(0.39)=4.9\times10^{-4} \simeq \eta_{\rm obs}. 
    \label{eq:etaDeltaV}
\end{equation}
Other examples of values are $\eta_\Delta(0.3)=8.8\times10^{-6}$,  $\eta_\Delta(0.35)=1.2\times10^{-4}$, $\eta_\Delta(0.4)=6.5\times10^{-4}$, $\eta_\Delta(0.45)=2.2\times10^{-3}$, and $\eta_\Delta(0.5)=5.3\times10^{-3}$. 

This section concludes that a plausible distribution of the magnetic field and the inclination of the magnetic axis to the spin axis of the WD-core merger remnant in the CD scenario for normal SNe Ia can account for the observed fraction of SNe Ia-CSM without the need to invoke a different scenario than that of most SNe Ia; the SNe Ia-CSM simply have short MED times. The magnetic field distribution is compatible with observations and studies of the merger process. Although I consider this very encouraging, the uncertainties in several parameters are very large: the exact magnetic field distribution, the inclination angle, and the initial rotation of the merger product. On top of these, the largest uncertainty is in the condition for explosion. For these, I will not study further the implications of equation (\ref{eq:etadeltaE}) beyond those for one scenario: SNe IA-CSM and SNIPs; the latter account for most normal SNe Ia.

% ==========================================
\section{Summary}
\label{sec:Summary}
% ==========================================

I analyzed the newly determined \citep{Desaietal2026} fraction of SNe Ia-CSM with ejecta-CSM interaction within $\simeq 0.3 \yr$ after explosion (equation \ref{eq:fc03}) together with the newly determined fraction of SNIPs \citep{Soker2025Tycho}. I compared the observed ratio of SNe Ia-CSM to SNIPs as given in equation (\ref{eq:etaobs}), $\eta_{\rm obs} \simeq 5 \times 10^{-4}$, with two theoretical crude estimates (note that SNe Ia-CSM are included in the group of SNIPs in the framework of the CD scenario).
In this study, I analyze the new estimate by \cite{Desaietal2026} of the above value of $\eta_{\rm obs}$. I note that \cite{Sharmaetal2023} find that SNe Ia-CSM comprise $\approx 0.02\%–0.2\%$ of all SNe Ia. They have some SNe Ia which were observed up to $\simeq 500$~day after the explosion, but their results are similar to those of \cite{Desaietal2026}. 
 
The first theoretical approach assumes that the explosion rate is constant up to a maximum time $t_{\rm SNIP}$ (equation \ref{eq:ProbabilityConstant}), as in some previous studies of the MED time distribution. For that assumption, the predicted ratio is $\eta_{\rm con} \approx 10^{-3}$ (equation \ref{eq:etaconstant}). This is a factor of two larger than the observed value. Given the simple assumptions and large uncertainties in the observationally determined ratio, I consider these two ratios compatible.    

The second theoretical approach, which I derived here, is to consider the angular-momentum loss from the merger to the explosion of the core-WD merger remnant. I took the MED time due to the angular momentum loss from \cite{IlkovSoker2012}, with somewhat different scaling (equation \ref{eq:taub}).
Although this expression contains more physics than the first theoretical approach, it has many large uncertainties. I considered a plausible distribution of the magnetic field and inclination angle between the magnetic and spin axes, based in part on observations (equation \ref{eq:Bdistribution2}). This gives an explosion rate (equation \ref{eq:Rate2}) that increases to a maximum value (at a MED time given by equation \ref{eq:tmax1}), and then slowly decreases, as Figure \ref{fig:Nmed} shows. I find that for a distribution width of the magnetic field of $\Delta \simeq 0.39$ the expected ratio $\eta_\Delta$ (equation \ref{eq:etadeltaE}) matches the observed value (equation \ref{eq:etaDeltaV}). This value of $\Delta$ gives a distribution that is crudely compatible with the expectation from observations and calculations as \cite{Briggsetal2018} give. The ratio is sensitive to the value of the width of the magnetic field distribution $\Delta$, ensuring that for other values of the parameters in equation (\ref{eq:taub}) the observed ratio can be matched with a reasonable value of $\Delta$.

This study's main conclusion is that reasonable distributions of MED times, i.e., the explosion rate as a function of time after core-WD merger, in the framework of the CD scenario account well for the ratio of SNe Ia having interaction with a CSM within a fraction of a year, SNe Ia-CSM, to the total number of SNe Ia that interact with a CSM within hundreds of years (SNIPs). According to the CD scenario, in both cases, the CSM is planetary nebula material, ejected by an AGB star and ionized by the central WD, which is here a merger product. This finding implies that the evolutionary pathway to produce SNe Ia-CSM is not rare, but rather the same evolutionary path that produces most SNe Ia, which I argue is the CD scenario: SNe Ia-CSM are SNIPs with WD merger products having a very short MED time, $\lesssim 300 \yr$. 

This study further emphasizes the important role of the magnetic field of the merger remnant in determining the MED time. In a recent study, \cite{Shiberetal2026} discussed the important role of magnetic fields in facilitating the deflagration-to-detonation transition during the explosion process in $M_{\rm Ch}$ explosions. It seems that the magnetic fields of the exploding WDs play a major role in normal SNe Ia.    

There are many open issues regarding the CD scenario, the main two are the simulations of the core-WD merger process to yield a WD remnant of $M_{\rm WD} \simeq M_{\rm Ch}$, and a full calculation of the MED time (Table \ref{Tab:Table1}). For that, I reiterate my view that, at this time, all scenarios should be considered; if not for normal SNe Ia, some can account for peculiar SNe Ia (e.g., \citealt{Soker2019Rev}). To start with, the classification of SN Ia scenarios should be transparent with respect to their properties and outcomes. The classification, limited to the SD and DD scenarios, does not do this. I propose the classification in Table \ref{Tab:Table1}.    

% ===============================
\section*{Acknowledgments}
% ===============================

I thank Adam Soker for his help with the calculations, and an anonymous referee for very detailed and helpful comments. I thank the Charles Wolfson Academic Chair at the Technion for supporting this research.

%%%%%%%%%%%%%%%%%%%%%%%%%%%

%%%%%%%%%%%%%%%%%%%%%%%%%%%

% %%%%%%%%%%%%  References %%%%%%%%%%%%%%%%%%%%%

%\newpage 


\begin{thebibliography}{}
\expandafter\ifx\csname natexlab\endcsname\relax\def\natexlab#1{#1}\fi
\providecommand{\url}[1]{\href{#1}{#1}}
\providecommand{\dodoi}[1]{doi:~\href{http://doi.org/#1}{\nolinkurl{#1}}}
\providecommand{\doeprint}[1]{\href{http://ascl.net/#1}{\nolinkurl{http://ascl.net/#1}}}
\providecommand{\doarXiv}[1]{\href{https://arxiv.org/abs/#1}{\nolinkurl{https://arxiv.org/abs/#1}}}

\bibitem[\protect\citeauthoryear{Ablimit}{2021}]{Ablimit2021} Ablimit I., 2021, PASP, 133, 074201. \dodoi{10.1088/1538-3873/ac025c}

\bibitem[\protect\citeauthoryear{Ablimit \& Maeda}{2019}]{AblimitMaeda2019} Ablimit I., Maeda K., 2019, ApJ, 871, 31. \dodoi{10.3847/1538-4357/aaf722} 

\bibitem[\protect\citeauthoryear{Aleo et al.}{2023}]{Aleoetal2023} Aleo P.~D., Malanchev K., Sharief S., Jones D.~O., Narayan G., Foley R.~J., Villar V.~A., et al., 2023, ApJS, 266, 9. \dodoi{10.3847/1538-4365/acbfba}

\bibitem[\protect\citeauthoryear{Benacquista et al.}{2003}]{Benacquistaetal2003} Benacquista M., Sedrakian D.~M., Hairapetyan M.~V., Shahabasyan K.~M., Sadoyan A.~A., 2003, ApJL, 596, L223. \dodoi{10.1086/379532}

\bibitem[\protect\citeauthoryear{Bhat et al.}{2026}]{Bhatetal2026} Bhat A., Pakmor R., Shen K.~J., Bauer E.~B., Rajamuthukumar A.~S., 2026, A\&A, 706, A375. \dodoi{10.1051/0004-6361/202557683} 

\bibitem[\protect\citeauthoryear{Boos, Townsley, \& Shen}{2024}]{Bossetal2024} Boos S.~J., Townsley D.~M., Shen K.~J., 2024, ApJ, 972, 200. \dodoi{10.3847/1538-4357/ad5da2}

\bibitem[\protect\citeauthoryear{Bora et al.}{2024}]{Boraetal2024} Bora Z., K{\"o}nyves-T{\' o}th R., Vink{\ 'o} J., B{\' a}nhidi D., B{\ '\i}r{\' o} I.~B., Bostroem K.~A., B{\' o}di A., et al., 2024, PASP, 136, 094201. \dodoi{10.1088/1538-3873/ad6e18}

\bibitem[\protect\citeauthoryear{Braudo \& Soker}{2024}]{BraudoSoker2024} Braudo J., Soker N., 2024, OJAp, 7, 7. \dodoi{10.21105/astro.2310.16554}

\bibitem[\protect\citeauthoryear{Braudo \& Soker}{2025}]{BraudoSoker2025RAA} Braudo J., Soker N., 2025, RAA, 25, 065012. \dodoi{10.1088/1674-4527/add567}

\bibitem[\protect\citeauthoryear{Bregman et al.}{2024}]{Bregmanetal2024} Bregman J.~N., Gnedin O.~Y., Seitzer P.~O., Qu Z., 2024, ApJL, 968, L6. \dodoi{10.3847/2041-8213/ad498f}

\bibitem[\protect\citeauthoryear{Briggs et al.}{2018}]{Briggsetal2018} Briggs G.~P., Ferrario L., Tout C.~A., Wickramasinghe D.~T., 2018, MNRAS, 478, 899. \dodoi{10.1093/mnras/sty1150} 

\bibitem[\protect\citeauthoryear{Burmester et al.}{2025}]{Burmesteretal2025} Burmester U.~P., Ferrario L., Pakmor R., Seitenzahl I.~R., 2025, MNRAS.tmp. \dodoi{10.1093/mnras/staf1128}


\bibitem[\protect\citeauthoryear{Callan et al.}{2025}]{Callanetal2025} Callan F.~P., Collins C.~E., Sim S.~A., Shingles L.~J., Pakmor R., Srivastav S., Pollin J.~M., et al., 2025, MNRAS, 539, 1404. \dodoi{10.1093/mnras/staf539}

\bibitem[\protect\citeauthoryear{Casabona \& Fisher}{2024}]{CasabonaFisher2024} Casabona G.~O., Fisher R.~T., 2024, ApJL, 962, L31. \dodoi{10.3847/2041-8213/ad2615}

\bibitem[\protect\citeauthoryear{Chen et al.}{2026}]{ChenCetal2026} Chen C., Sun N.-C., Xi Q., Tinyanont S., Aguado D., P{\'e}rez-Fournon I., Poidevin F., et al., 2026, ApJ, 997, 72. \dodoi{10.3847/1538-4357/ae285c} 


\bibitem[\protect\citeauthoryear{Contopoulos \& Spitkovsky}{2006}]{ContopoulosSpitkovsky2006} Contopoulos I., Spitkovsky A., 2006, ApJ, 643, 1139. \dodoi{10.1086/501161} 

\bibitem[\protect\citeauthoryear{Court et al.}{2026}]{Courtetal2026} Court T., Badenes C., Lee S.-H., Patnaude D., Bravo E., 2026, ApJ, 996, 100. \dodoi{10.3847/1538-4357/ae1a7e} 


\bibitem[\protect\citeauthoryear{Cui et al.}{2022}]{CuiMeng2022} Cui Y., Meng X., Podsiadlowski P., Song R., 2022, A\&A, 667, A154. \dodoi{10.1051/0004-6361/202141335}

\bibitem[\protect\citeauthoryear{Das et al.}{2026}]{Dasetal2026} Das P., Seitenzahl I.~R., Ghavamian P., Ruiter A.~J., Laming J.~M., Murphy S.~J., O'Donnel C., 2026, MNRAS, 548, 1. \dodoi{10.1093/mnras/stag596}

\bibitem[\protect\citeauthoryear{Das et al.}{2025}]{Dasetal2025NatAs} Das P., Seitenzahl I.~R., Ruiter A.~J., R{\"o}pke F.~K., Pakmor R., Vogt F.~P.~A., Collins C.~E., et al., 2025, NatAs, 9, 1356. \dodoi{10.1038/s41550-025-02589-5}


\bibitem[\protect\citeauthoryear{DerKacy et al.}{2024}]{DerKacyetal2024} DerKacy J.~M., Ashall C., Hoeflich P., Baron E., Shahbandeh M., Shappee B.~J., Andrews J., et al., 2024, ApJ, 961, 187. \dodoi{10.3847/1538-4357/ad0b7b}

\bibitem[\protect\citeauthoryear{Desai et al.}{2026}]{Desaietal2026} Desai D.~D., Shappee B.~J., Kochanek C.~S., Stanek K.~Z., Ashall C., Beacom J.~F., Burns C.~R., et al., 2026, arXiv, arXiv:2602.00223. \dodoi{10.48550/arXiv.2602.00223} 


\bibitem[\protect\citeauthoryear{Dickel \& Jones}{1985}]{DickelJones1985} Dickel J.~R., Jones E.~M., 1985, ApJ, 288, 707. \dodoi{10.1086/162837}

\bibitem[\protect\citeauthoryear{Di Stefano, Voss, \& Claeys}{2011}]{DiStefanoetal2011} Di Stefano R., Voss R., Claeys J.~S.~W., 2011, ApJL, 738, L1. \dodoi{10.1088/2041-8205/738/1/L1}

\bibitem[\protect\citeauthoryear{Ferrand et al.}{2025}]{Ferrandetal2025} Ferrand G., Pakmor R., Fujimaru Y., Lee S.-H., Safi-Harb S., Nagataki S., R{\"o}pke F.~K., et al., 2025, ApJ, 995, 85. \dodoi{10.3847/1538-4357/ae1602} 

\bibitem[\protect\citeauthoryear{Gaba et al.}{2025}]{Gabaetal2025NewA} Gaba J., Thakur R.~K., Sharma N., Verma D., Gupta S., 2025, NewA, 120, 102411. \dodoi{10.1016/j.newast.2025.102411}

\bibitem[\protect\citeauthoryear{Gaba et al.}{2026}]{Gabaetal2026} Gaba J., Thakur R.~K., Verma D., Sharma N., Gupta S., 2026, arXiv, arXiv:2603.02867
\dodoi{10.48550/arXiv.2603.02867} 


\bibitem[\protect\citeauthoryear{Garc{\'\i}a-Berro et al.}{2012}]{GarciaBerroetal2012} Garc{\'\i}a-Berro E., Lor{\'e}n-Aguilar P., Aznar-Sigu{\'a}n G., Torres S., Camacho J., Althaus L.~G., C{\'o}rsico A.~H., et al., 2012, ApJ, 749, 25. \dodoi{10.1088/0004-637X/749/1/25}

\bibitem[\protect\citeauthoryear{Glanz et al.}{2025a}]{Glanzetal2025NatAs} Glanz H., Perets H.~B., Bhat A., Pakmor R., 2025a, NatAs, 9, 1523. \dodoi{10.1038/s41550-025-02633-4} 

\bibitem[\protect\citeauthoryear{Glanz, Perets, \& Bhat}{2026}]{Glanzetal2026} Glanz H., Perets H.~B., Bhat A., 2026, arXiv, arXiv:2510.03396. \dodoi{10.48550/arXiv.2510.03396}

\bibitem[\protect\citeauthoryear{Glanz, Perets, \& Pakmor}{2025b}]{Glanzetal2025b} Glanz H., Perets H.~B., Pakmor R., 2025b, ApJ, 988, 184. \dodoi{10.3847/1538-4357/ade146}

\bibitem[\protect\citeauthoryear{Griffith et al.}{2025}]{Griffithetal2025} Griffith O., Showerman G., Sarbadhicary S.~K., Harris C.~E., Chomiuk L., Sollerman J., Lundqvist P., et al., 2025, ApJ, 995, 54. \dodoi{10.3847/1538-4357/ae17b0} 

\bibitem[\protect\citeauthoryear{Guo et al.}{2025}]{GuoMeng2025RAA} Guo B., Meng X., Tian Z., Luo J., Liu Z., 2025, RAA, 25, 015018. \dodoi{10.1088/1674-4527/ada2e9}

\bibitem[\protect\citeauthoryear{Heiger et al.}{2026}]{Heigeretal2026} Heiger M.~E., Ji A.~P., Speagle J.~S., Li T.~S., Savino A., Sandford N.~R., Kirby E.~N., et al., 2026, arXiv, arXiv:2602.22333. \dodoi{10.48550/arXiv.2602.22333}

\bibitem[\protect\citeauthoryear{Hillman, Michaelis, \& Perets}{2026}]{Hillmanetal2026} Hillman Y., Michaelis A., Perets H.~B., 2026, JHEAp, 53, 100605. \dodoi{10.1016/j.jheap.2026.100605}
 
 
\bibitem[\protect\citeauthoryear{Hillman et al.}{2020}]{Hillmanetal2020} Hillman Y., Shara M., Prialnik D., Kovetz A., 2020, AdSpR, 66, 1072. \dodoi{10.1016/j.asr.2019.08.029} 

\bibitem[\protect\citeauthoryear{Hoeflich}{2017}]{Hoeflich2017} Hoeflich P., 2017, in Handbook of Supernovae, Springer International Publishing AG, 2017, p. 1151 \dodoi{10.1007/978-3-319-21846-5\_56}


\bibitem[\protect\citeauthoryear{Hoeflich et al.}{2025}]{Hoeflichetal2026} Hoeflich P., Fereidouni E., Fisher A., Mera T., Ashall C., Brown P., Baron E., et al., 2025, arXiv, arXiv:2501.07654. \dodoi{10.48550/arXiv.2501.07654} 

\bibitem[\protect\citeauthoryear{Hoogendam et al.}{2025a}]{Hoogendametal2025a} Hoogendam W.~B., Ashall C., Jones D.~O., Shappee B.~J., Tucker M.~A., Huber M.~E., Auchettl K., et al., 2025a, ApJ, 988, 209. \dodoi{10.3847/1538-4357/ade787}

\bibitem[\protect\citeauthoryear{Hoogendam et al.}{2025b}]{Hoogendametal2025b} Hoogendam W.~B., Jones D.~O., Ashall C., Shappee B.~J., Foley R.~J., Tucker M.~A., Huber M.~E., et al., 2025b, OJAp, 8, 120. \dodoi{10.33232/001c.143462} 

\bibitem[\protect\citeauthoryear{Igoshev, Perets, \& Hallakoun}{2023}]{Igoshevetal2023} Igoshev A.~P., Perets H., Hallakoun N., 2023, MNRAS, 518, 6223. \dodoi{10.1093/mnras/stac3488}

\bibitem[\protect\citeauthoryear{Ilkov \& Soker}{2012}]{IlkovSoker2012} Ilkov M., Soker N., 2012, MNRAS, 419, 1695. \dodoi{10.1111/j.1365-2966.2011.19833.x} 

\bibitem[\protect\citeauthoryear{Ito et al.}{2025}]{Itoetal2025} Ito D., Sano H., Nakazawa K., Mitsuishi I., Fukui Y., Sudou H., Takaba H., 2025, ApJ, 978, 123. \dodoi{10.3847/1538-4357/ad95f5}

\bibitem[\protect\citeauthoryear{Iwata \& Maeda}{2025}]{IwataMaeda2025} Iwata K., Maeda K., 2025, ApJ, 987, 21. \dodoi{10.3847/1538-4357/add5ed}

\bibitem[\protect\citeauthoryear{Jha, Maguire, \& Sullivan}{2019}]{Jhaetal2019NatAs} Jha S.~W., Maguire K., Sullivan M., 2019, NatAs, 3, 706. \dodoi{10.1038/s41550-019-0858-0}

\bibitem[\protect\citeauthoryear{Joshi, Strolger, \& Zenati}{2024}]{Joshietal2024} Joshi B.~A., Strolger L.-G., Zenati Y., 2024, ApJ, 974, 15. \dodoi{10.3847/1538-4357/ad6843}

\bibitem[\protect\citeauthoryear{Justham}{2011}]{Justham2011} Justham S., 2011, ApJL, 730, L34. \dodoi{10.1088/2041-8205/730/2/L34}

\bibitem[\protect\citeauthoryear{Kashi \& Soker}{2011}]{KashiSoker2011} Kashi A., Soker N., 2011, MNRAS, 417, 1466. \dodoi{10.1111/j.1365-2966.2011.19361.x} 

\bibitem[\protect\citeauthoryear{Kepler et al.}{2013}]{Kepleretal2013} Kepler S.~O., Pelisoli I., Jordan S., Kleinman S.~J., Koester D., K{\"u}lebi B., Pe{\c{c}}anha V., et al., 2013, MNRAS, 429, 2934. \dodoi{10.1093/mnras/sts522} 

\bibitem[\protect\citeauthoryear{Kerzendorf et al.}{2018}]{Kerzendorfetal2018} Kerzendorf W.~E., Long K.~S., Winkler P.~F., Do T., 2018, MNRAS, 479, 5696. \dodoi{10.1093/mnras/sty1863}

\bibitem[\protect\citeauthoryear{Kerzendorf et al.}{2013}]{Kerzendorfetal2013} Kerzendorf W.~E., Yong D., Schmidt B.~P., Simon J.~D., Jeffery C.~S., Anderson J., Podsiadlowski P., et al., 2013, ApJ, 774, 99. \dodoi{10.1088/0004-637X/774/2/99}

\bibitem[\protect\citeauthoryear{Ko et al.}{2024}]{Koetal2024} Ko T., Suzuki H., Kashiyama K., Uchida H., Tanaka T., Tsuna D., Fujisawa K., et al., 2024, ApJ, 969, 116. \dodoi{10.3847/1538-4357/ad4d99}

\bibitem[\protect\citeauthoryear{Kobashi et al.}{2024}]{Kobashietal2024} Kobashi R., Lee S.-H., Tanaka T., Maeda K., 2024, ApJ, 961, 32. \dodoi{10.3847/1538-4357/ad05c2}

\bibitem[\protect\citeauthoryear{Kumar, Prust, \& Bildsten}{2025}]{Kumaretal2025} Kumar G., Prust L.~J., Bildsten L., 2025, ApJ, 992, 2. \dodoi{:10.3847/1538-4357/adfdd7}

\bibitem[\protect\citeauthoryear{Kwok et al.}{2025}]{KwokLetal2026} Kwok L.~A., Liu C., Jha S.~W., Blondin S., Larison C., Miller A.~A., Dai M., et al., 2025, arXiv, arXiv:2510.09760. \dodoi{10.48550/arXiv.2510.09760}

\bibitem[\protect\citeauthoryear{Li, Liu, \& Wang}{2023}]{Lietal2023RAA} Li L.-H., Liu D.-D., Wang B., 2023, RAA, 23, 075010. \dodoi{10.1088/1674-4527/acd0ea}

\bibitem[\protect\citeauthoryear{Lim et al.}{2024}]{Limetal2024} Lim G., Im M., Paek G.~S.~H., Yoon S.-C., Imsng Team, 2024, ASPC, 536, 29

\bibitem[\protect\citeauthoryear{Liu, R{\" o}pke, \& Han}{2023}]{Liuetal2023Rev} Liu Z.-W., R{\" o}pke F.~K., Han Z., 2023, RAA, 23, 082001. \dodoi{10.1088/1674-4527/acd89e}

\bibitem[\protect\citeauthoryear{Livio \& Mazzali}{2018}]{LivioMazzali2018} Livio M., Mazzali P., 2018, PhR, 736, 1. \dodoi{10.1016/j.physrep.2018.02.002}

\bibitem[\protect\citeauthoryear{Macrie et al.}{2026}]{Macrieetal2026} Macrie C.~W., Larison C., Sears H., Kwok L.~A., Jha S.~W., Dai M., Johansson J., et al., 2026, arXiv, arXiv:2604.09777

\bibitem[\protect\citeauthoryear{Maeda \& Terada}{2016}]{MaedaTerada2016} Maeda K., Terada Y., 2016, IJMPD, 25, 1630024. \dodoi{10.1142/S021827181630024X}

\bibitem[\protect\citeauthoryear{Magee et al.}{2025}]{Mageeetal2025} Magee M.~R., Killestein T.~L., Pursiainen M., Godson B., Jarvis D., Jim{\'e}nez-Palau C., Lyman J.~D., et al., 2025, MNRAS, 543, 3731. \dodoi{10.1093/mnras/staf1675} 


\bibitem[\protect\citeauthoryear{Malheiro et al.}{2026}]{Malheiroetal2026} Malheiro M., Borges S.~V., Coelho J.~G., Kianfar K., Lobato R.~V., Otoniel E., Rueda J.~A., et al., 2026, JHEAp, 53, 100593. \dodoi{10.1016/j.jheap.2026.100593} 

\bibitem[\protect\citeauthoryear{Mandal et al.}{2024}]{Mandaletal2024} Mandal S., Duffell P.~C., Polin A., Milisavljevic D., 2024, ApJ, 972, 87. \dodoi{10.3847/1538-4357/ad5daa}

\bibitem[\protect\citeauthoryear{Mandal et al.}{2025a}]{Mandaletal2025SNR0509} Mandal S., Ghavamian P., Das P., Seitenzahl I.~R., Mohamed S., Ruiter A.~J., 2025a, arXiv, arXiv:2509.02422. \dodoi{10.48550/arXiv.2509.02422}

\bibitem[\protect\citeauthoryear{Mandal et al.}{2025b}]{MandalSetal2025Tycho} Mandal S., Torres-Alb{\ 'a} N., Badenes C., Mohamed S., 2025b, arXiv, arXiv:2508.10752. \dodoi{10.48550/arXiv.2508.10752}

\bibitem[\protect\citeauthoryear{Maoz, Mannucci, \& Nelemans}{2014}]{Maozetal2014} Maoz D., Mannucci F., Nelemans G., 2014, ARA\&A, 52, 107. \dodoi{10.1146/annurev-astro-082812-141031} 


\bibitem[\protect\citeauthoryear{Mehta et al.}{2024}]{Mehtaetal2024} Mehta V., Sullivan J., Fisher R., Ohshiro Y., Yamaguchi H., Bhargava K., Neopane S., 2024, MNRAS, 532, 1087. \dodoi{10.1093/mnras/stae1559}


\bibitem[\protect\citeauthoryear{Mehta et al.}{2026}]{Mehtaetal2026} Mehta V., Tiwari V., Pakmor R., Singh D., Fisher R., 2026, arXiv, arXiv:2602.23414. \dodoi{10.48550/arXiv.2602.23414} 

\bibitem[\protect\citeauthoryear{Meng \& Podsiadlowski}{2017}]{MengPodsiadlowski2017} Meng X., Podsiadlowski P., 2017, MNRAS, 469, 4763. \dodoi{10.1093/mnras/stx1137}


\bibitem[\protect\citeauthoryear{Michaelis, Hillman, \& Perets}{2025}]{Michaelisetal2025} Michaelis A., Hillman Y., Perets H.~B., 2025, arXiv, arXiv:2510.20904. \dodoi{10.48550/arXiv.2510.20904}

\bibitem[\protect\citeauthoryear{Michaelis \& Perets}{2025}]{MichaelisPerets2025} Michaelis A., Perets H.~B., 2025, arXiv, arXiv:2507.16907
\dodoi{10.48550/arXiv.2507.16907}

\bibitem[\protect\citeauthoryear{Mor{\'a}n-Fraile et al.}{2024}]{MoranFraileetal2024} Mor{\'a}n-Fraile J., Holas A., R{\"o}pke F.~K., Pakmor R., Schneider F.~R.~N., 2024, A\&A, 683, A44. \dodoi{10.1051/0004-6361/202347769}

\bibitem[\protect\citeauthoryear{Neopane et al.}{2022}]{Neopaneetal2022} Neopane S., Bhargava K., Fisher R., Ferrari M., Yoshida S., Toonen S., Bravo E., 2022, ApJ, 925, 92. \dodoi{10.3847/1538-4357/ac3b52}

\bibitem[\protect\citeauthoryear{O'Hora et al.}{2025}]{OHoraetal2025} O'Hora J., Ashall C., Shahbandeh M., Hsiao E., Hoeflich P., Stritzinger M.~D., Galbany L., et al., 2025, ApJ, 984, 34. \dodoi{10.3847/1538-4357/adc37c}

\bibitem[\protect\citeauthoryear{Padilla Gonzalez et al.}{2024}]{PadillaGonzalezetal2024} Padilla Gonzalez E., Howell D.~A., Terreran G., McCully C., Newsome M., Burke J., Farah J., et al., 2024, ApJ, 964, 196. \dodoi{10.3847/1538-4357/ad19c9}

\bibitem[\protect\citeauthoryear{Pakmor et al.}{2011}]{Pakmoretal2011} Pakmor R., Hachinger S., R{\"o}pke F.~K., Hillebrandt W., 2011, A\&A, 528, A117. \dodoi{10.1051/0004-6361/201015653} 

\bibitem[\protect\citeauthoryear{Pakmor et al.}{2026}]{Pakmoretal2026} Pakmor R., Shen K.~J., Bhat A., Rajamuthukumar A.~S., Collins C.~E., O'Donnell C., Bauer E.~B., et al., 2026, A\&A, 706, A239. \dodoi{10.1051/0004-6361/202557670}


\bibitem[\protect\citeauthoryear{Papish et al.}{2015}]{Papishetal2015} Papish O., Soker N., Garc{\ '\i}a-Berro E., Aznar-Sigu{\' a}n G., 2015, MNRAS, 449, 942. \dodoi{10.1093/mnras/stv337}

\bibitem[\protect\citeauthoryear{Palicio et al.}{2024}]{Palicioetal2024} Palicio P.~A., Matteucci F., Della Valle M., Spitoni E., 2024, A\&A, 689, A203. \dodoi{10.1051/0004-6361/202449740}

\bibitem[\protect\citeauthoryear{Pan, Ruiz-Lapuente, \& Gonz{\'a}lez Hern{\'a}ndez}{2026}]{Panetal2026} Pan K.-C., Ruiz-Lapuente P., Gonz{\'a}lez Hern{\'a}ndez J.~I., 2026, ApJ, 996, 35. \dodoi{10.3847/1538-4357/ae1b9e} 

\bibitem[\protect\citeauthoryear{Patel et al.}{2026}]{Pateletal2026} Patel K., Dongre A., Fisher R., Poludnenko A., Gamezo V., Ugalino M., Byrohl C., 2026, arXiv, arXiv:2605.21575. \dodoi{10.48550/arXiv.2605.21575}


\bibitem[\protect\citeauthoryear{Pearson et al.}{2024}]{Pearsonetal2024} Pearson J., Sand D.~J., Lundqvist P., Galbany L., Andrews J.~E., Bostroem K.~A., Dong Y., et al., 2024, ApJ, 960, 29. \dodoi{10.3847/1538-4357/ad0153}

\bibitem[\protect\citeauthoryear{Perets et al.}{2019}]{Peretsetal2019} Perets H.~B., Zenati Y., Toonen S., Bobrick A., 2019, arXiv:1910.07532. \dodoi{10.48550/arXiv.1910.07532}


\bibitem[\protect\citeauthoryear{Phillips et al.}{2024}]{Phillipsetal2024} Phillips M.~M., Ashall C., Brown P.~J., Galbany L., Tucker M.~A., Burns C.~R., Contreras C., et al., 2024, ApJS, 273, 16. \dodoi{10.3847/1538-4365/ad4f7e}

\bibitem[\protect\citeauthoryear{Pollin et al.}{2024}]{Polinetal2024} Pollin J.~M., Sim S.~A., Pakmor R., Callan F.~P., Collins C.~E., Shingles L.~J., R{\" o}pke F.~K., et al., 2024, MNRAS, 533, 3036. \dodoi{10.1093/mnras/stae1909}

\bibitem[\protect\citeauthoryear{Pollin et al.}{2026}]{Pollinetal2026}  Pollin J.~M., Sim S.~A., Shingles L.~J., Pakmor R., Callan F.~P., Collins C.~E., R{\"o}pke F.~K., et al., 2026, MNRAS.tmp. \dodoi{10.1093/mnras/stag735} 

\bibitem[\protect\citeauthoryear{Rajavel, Townsley, \& Shen}{2025}]{Rajaveletal2025} Rajavel N., Townsley D.~M., Shen K.~J., 2025, ApJ, 979, 54. \dodoi{10.3847/1538-4357/ada034}

\bibitem[\protect\citeauthoryear{Reynolds}{1985}]{Reynolds1985} Reynolds R.~J., 1985, ApJ, 288, 622. dodoi{10.1086/162828}

\bibitem[\protect\citeauthoryear{Ruiter}{2020}]{Ruiter2020} Ruiter A.~J., 2020, IAUS, 357, 1. \dodoi{10.1017/S1743921320000587}

\bibitem[\protect\citeauthoryear{Ruiter \&  Seitenzahl}{2025}]{RuiterSeitenzahl2025} Ruiter A.~J., Seitenzahl I.~R., 2025, A\&ARv, 33, 1. \dodoi{10.1007/s00159-024-00158-9}

\bibitem[\protect\citeauthoryear{Ruiz-Lapuente}{2019}]{RuizLapuente2019} Ruiz-Lapuente P., 2019, NewAR, 85, 101523. \dodoi{10.1016/j.newar.2019.101523} 

\bibitem[\protect\citeauthoryear{Schaefer}{2025a}]{Schaefer2025a} Schaefer B.~E., 2025a, ApJ, 980, 156. \dodoi{10.3847/1538-4357/ada38d} 

\bibitem[\protect\citeauthoryear{Schaefer}{2025b}]{Schaefer2025b} Schaefer B.~E., 2025b, ApJ, 991, 111. \dodoi{10.3847/1538-4357/adfa16}

\bibitem[\protect\citeauthoryear{Schaefer}{2025c}]{Schaefer2025c} Schaefer B.~E., 2025c, ApJ, 994, 187. \dodoi{10.3847/1538-4357/ae13e4}

\bibitem[\protect\citeauthoryear{Schaefer}{2026}]{Schaefer2026} Schaefer B.~E., 2026, ApJ, 1000, 219. \dodoi{10.3847/1538-4357/ae3aa2} 

\bibitem[\protect\citeauthoryear{Schaefer \& Myers}{2025}]{SchaeferMyers2025} Schaefer B.~E., Myers G., 2025, ApJ, 991, 110. \dodoi{10.3847/1538-4357/adf961} 

\bibitem[\protect\citeauthoryear{Schaefer et al.}{2026}]{Schaeferetal2026} Schaefer B.~E., Pearce A., Love T., Shara M.~M., Townsend L., Murphy S.~J., Corbally C.~J., 2026, ApJ, 1000, 125. \dodoi{10.3847/1538-4357/ae4356} 


\bibitem[\protect\citeauthoryear{Schinasi-Lemberg \& Kushnir}{2025}]{SchinasiLembergKushnir2025} Schinasi-Lemberg E., Kushnir D., 2025, MNRAS, 536, 3041. \dodoi{10.1093/mnras/stae2770}

\bibitem[\protect\citeauthoryear{Sharma et al.}{2023}]{Sharmaetal2023} Sharma Y., Sollerman J., Fremling C., Kulkarni S.~R., De K., Irani I., Schulze S., et al., 2023, ApJ, 948, 52. \dodoi{10.3847/1538-4357/acbc16}

\bibitem[\protect\citeauthoryear{Sharon \& Kushnir}{2025}]{sharonKushnirWygoda2025}  Sharon A., Kushnir D., Wygoda N., 2025, MNRAS, 540, 3247. \dodoi{10.1093/mnras/staf808}


\bibitem[\protect\citeauthoryear{Sharon, Kushnir, \& Schinasi-Lemberg}{2025}]{Sharonetal2025} Sharon A., Kushnir D., Schinasi-Lemberg E., 2025, MNRAS, 540, 348. \dodoi{10.1093/mnras/stae2450}


\bibitem[\protect\citeauthoryear{Shen, Boos, \& Townsley}{2024}]{Shenetal2024} Shen K.~J., Boos S.~J., Townsley D.~M., 2024, ApJ, 975, 127. \dodoi{10.3847/1538-4357/ad7379}

\bibitem[\protect\citeauthoryear{Shiber et al.}{2026}]{Shiberetal2026} Shiber S., Hoeflich P., Mera T., Fereidouni E., Levy Z., Maci D., Ashall C., et al., 2026, arXiv, arXiv:2602.11341. \dodoi{10.48550/arXiv.2602.11341} 

\bibitem[\protect\citeauthoryear{Simotas, Bildsten, \& Prust}{2025}]{Simotasetal2025} Simotas K., Bildsten L., Prust L.~J., 2025, ApJ, 994, 23. \dodoi{10.3847/1538-4357/ae0734}


\bibitem[\protect\citeauthoryear{Soker}{2018}]{Soker2018Rev} Soker N., 2018, SCPMA, 61, 49502. \dodoi{10.1007/s11433-017-9144-4}

\bibitem[\protect\citeauthoryear{Soker}{2019a}]{Soker2019Rev} Soker N., 2019a, NewAR, 87, 101535. \dodoi{10.1016/j.newar.2020.101535}

\bibitem[\protect\citeauthoryear{Soker}{2019b}]{Soker2019CEEDTD} Soker N., 2019b, MNRAS, 490, 2430. \dodoi{10.1093/mnras/stz2817}

\bibitem[\protect\citeauthoryear{Soker}{2022}]{Soker2022RAA} Soker N., 2022, RAA, 22, 035025. \dodoi{10.1088/1674-4527/ac4d25}

\bibitem[\protect\citeauthoryear{Soker}{2024a}]{Soker2024Rev} Soker N., 2024a, OJAp, 7, 31. \dodoi{10.33232/001c.117147}

\bibitem[\protect\citeauthoryear{Soker}{2024b}]{Soker2024RAAPN} Soker N., 2024b, RAA, 24, 015012. \dodoi{10.1088/1674-4527/ad0ded}


\bibitem[\protect\citeauthoryear{Soker}{2024c}]{Soker2024Comment} Soker N., 2024c, arXiv:2406.07363. \dodoi{10.48550/arXiv.2406.07363}

\bibitem[\protect\citeauthoryear{Soker}{2025d}]{Soker2025SNR0509} Soker N., 2025d, OJAp, 8, 36. \dodoi{10.33232/001c.134113}

\bibitem[\protect\citeauthoryear{Soker}{2025e}]{Soker2025DDMED} Soker N., 2025e, RAA, 25, 111001.. \dodoi{10.1088/1674-4527/ae0703}


\bibitem[\protect\citeauthoryear{Soker}{2025f}]{Soker2025Tycho} Soker N., 2025f, Univ, 11, 377. \dodoi{10.3390/universe11110377} 

\bibitem[\protect\citeauthoryear{Toonen, Perets, \& Hamers}{2018}]{Toonenetal2018} Toonen S., Perets H.~B., Hamers A.~S., 2018, A\&A, 610, A22. \dodoi{10.1051/0004-6361/201731874}

\bibitem[\protect\citeauthoryear{Tsalapatas et al.}{2025}]{Tsalapatasetal2025} Tsalapatas K., Sollerman J., Chiba R., Kool E., Johansson J., Rosswog S., Schulze S., et al., 2025, A\&A, 704, A135. \dodoi{10.1051/0004-6361/202556369}

\bibitem[\protect\citeauthoryear{Tsebrenko \& Soker}{2013}]{TsebrenkoSoker2013} Tsebrenko D., Soker N., 2013, MNRAS, 435, 320. \dodoi{10.1093/mnras/stt1301}

\bibitem[\protect\citeauthoryear{Tsebrenko \& Soker}{2015}]{TsebrenkoSoker2015} Tsebrenko D., Soker N., 2015, MNRAS, 447, 2568. \dodoi{10.1093/mnras/stu2567}

\bibitem[\protect\citeauthoryear{Tweedy, Martos, \& Noriega-Crespo}{1995}]{Tweedyetal1995} Tweedy R.~W., Martos M.~A., Noriega-Crespo A., 1995, ApJ, 447, 257. dodoi{10.1086/175871}

\bibitem[\protect\citeauthoryear{Uchida et al.}{2024}]{Uchidaetal2024} Uchida H., Kasuga T., Maeda K., Lee S.-H., Tanaka T., Bamba A., 2024, ApJ, 962, 159. \dodoi{10.3847/1538-4357/ad1ff3}


\bibitem[\protect\citeauthoryear{Vathachira, Hillman, \& Kashi}{2024}]{Vathachiraetal2024} Vathachira I.~B., Hillman Y., Kashi A., 2024, MNRAS, 527, 4806. \dodoi{10.1093/mnras/stad3507}

\bibitem[\protect\citeauthoryear{Vathachira, Hillman, \& Kashi}{2026}]{Vathachiraetal2026} Vathachira I.~B., Hillman Y., Kashi A., 2026, ApJ, 997, 278. \dodoi{10.3847/1538-4357/ae27c7} 


\bibitem[\protect\citeauthoryear{Vinko, Szalai, \& K{\" o}nyves-T{\' o}th}{2023}]{Vinkoetal2023} Vinko J., Szalai T., K{\" o}nyves-T{\' o}th R., 2023, Univ, 9, 244. \dodoi{10.3390/universe9060244}

\bibitem[\protect\citeauthoryear{Wang}{2018}]{Wang2018} Wang B., 2018, RAA, 18, 049. \dodoi{10.1088/1674-4527/18/5/49}
  
\bibitem[\protect\citeauthoryear{Wang et al.}{2024}]{Wangetal2024} Wang Q., Rest A., Dimitriadis G., Ridden-Harper R., Siebert M.~R., Magee M., Angus C.~R., et al., 2024, ApJ, 962, 17. \dodoi{10.3847/1538-4357/ad0edb}


\bibitem[\protect\citeauthoryear{Wang \& Meng}{2025}]{WangMeng2025} Wang X., Meng X., 2025, A\&A, 699, A35. \dodoi{10.1051/0004-6361/202554449}

\bibitem[\protect\citeauthoryear{Wang, Chen, \& Pan}{2025}]{WangChenPan2025} Wang Y.-H., Chen H.-P., Pan K.-C., 2025, ApJ, 989, 72. \dodoi{10.3847/1538-4357/adeb71}

\bibitem[\protect\citeauthoryear{Wu et al.}{2025}]{Wuetal2025} Wu W., Jiang J.-. an ., Meng D., Xu Z., Maeda K., Doi M., Nomoto K., et al., 2025, ApJ, 991, 148. \dodoi{:10.3847/1538-4357/adf05a}
 
\bibitem[\protect\citeauthoryear{Yoon \& Langer}{2005}]{YoonLanger2005} Yoon S.-C., Langer N., 2005, A\&A, 435, 967. \dodoi{10.1051/0004-6361:20042542}

\bibitem[\protect\citeauthoryear{Zenati et al.}{2023}]{Zenatietal2023} Zenati Y., Perets H.~B., Dessart L., Jacobson-Gal{\ 'a}n W.~V., Toonen S., Rest A., 2023, ApJ, 944, 22. \dodoi{10.3847/1538-4357/acaf65}

\bibitem[\protect\citeauthoryear{Zenati, Toonen, \& Perets}{2019}]{Zenatietal2019} Zenati Y., Toonen S., Perets H.~B., 2019, MNRAS, 482, 1135. \dodoi{10.1093/mnras/sty2723}

\bibitem[\protect\citeauthoryear{Zhao, Maeda, \& Wang}{2026}]{ZhaoXetal2026}  Zhao X., Maeda K., Wang X., 2026, RAA, 26, 065002. \dodoi{10.1088/1674-4527/ae4baf} 


\bibitem[\protect\citeauthoryear{Zingale et al.}{2024}]{Zingaleetal2024} Zingale M., Chen Z., Rasmussen M., Polin A., Katz M., Smith Clark A., Johnson E.~T., 2024, ApJ, 966, 150. \dodoi{10.3847/1538-4357/ad3441}


\end{thebibliography}
\end{document}